\documentclass[aps,prb,twocolumn,showpacs,superscriptaddress,preprintnumbers]{revtex4-2}
\usepackage{graphicx} 
\usepackage{amsmath}
\usepackage[T1]{fontenc}
\usepackage{graphicx,epstopdf}
\usepackage{graphicx}
\usepackage{epstopdf}
\usepackage{mathtools}
\usepackage{tikz}
\usepackage{color}
\usetikzlibrary{shapes.misc,shadows}

\newcommand{\be}{\begin{equation}}
\newcommand{\ee}{\end{equation}}
\newcommand{\bea}{\begin{eqnarray}}
\newcommand{\eea}{\end{eqnarray}}
\newcommand{\bse}{\begin{subequations}}
\newcommand{\ese}{\end{subequations}}

\begin{document}
\title{Accuracy and speed of elongation in a minimal model of DNA replication}
\author{M Sahoo}
\email{jolly.iopb@gmail.com}
\affiliation{Department of Physics, University of Kerala, Kariavattom Campus-6955881, India}
\affiliation{School of Physics, Indian Institute of Science Education and Research, Thiruvananthapuram-695551, India}
\author{Arsha N}
\affiliation{Department of Physics, University of Kerala, Kariavattom Campus-6955881, India}
\author{P R Baral}
\affiliation{School of Physics, Indian Institute of Science Education and Research, Thiruvananthapuram-695551, India}
\author{S Klumpp}
\email{stefan.klumpp@phys.uni-goettingen.de}
\affiliation{Institute for the Dynamics of Complex Systems, University of G\"ottingen, G\"ottingen, Germany}
\date{\today}

\begin{abstract}
Being a dual purpose enzyme, the DNA polymerase is responsible for elongation of the newly formed DNA strand as well as cleaving the erroneous growth in case of a misincorporation. The efficiency of replication depends on the coordination of the polymerization and exonuclease activity of DNA polymerase. 
Here we propose and analyze a minimal kinetic model of DNA replication and determine exact expressions for the velocity of elongation and the accuracy of replication. We first analyze the case without exonuclease activity. In that case, accuracy is determined by a kinetic competition between stepping and unbinding, with discrimination between correct and incorrect nucleotides in both transitions. We then include exonuclease activity and ask how different modes of additional discrimination in the exonuclease pathway can improve the accuracy while limiting the detrimental effect of exonuclase on the speed of replication. In this way, we ask  how the kinetic parameters of the model have to be set to coordinate the two activities of the enzyme for high accuracy and high speed. The analysis also shows that the design of a replication system does not universally have to follow the speed-accuracy trade-off rule, although it does in the biologically realized parameter range. The accuracy of the process is mainly controlled by the crucial role of stepping after erroneous incorporation, which has impact on both polymerase and exonuclease  activities of DNA polymerase.
\end{abstract}
\maketitle

\section{\textbf{Introduction}}
The genetic information of a cell or an organism is stored in its DNA (deoxyribonucleic acid) and transmitted to the next generation through the process of DNA replication ~\cite{benkovic2001replisome,kornberg1992dna}. DNA replication is a complex process that involves multiple enzymes including helicases,  nucleases and topoisomerases  on\cite{hopfield1974kinetic,yager1991thermodynamic,tubulekas1993suppression,
wohlgemuth2011evolutionary,van2007stochastic}. At its core, however, is DNA polymerase (DNAP), which catalyzes the actual copying of the genetic information
\cite{patel1991pre}. During the replication process, the DNAP acts as a molecular motor and moves step by step on the template DNA strand converting  chemical energy into directed motion \cite{howard2001mechanics,kolomeisky2007molecular,schliwa2003molecular}. At the same time it copies the genetic information of the DNA sequence along which it moves into a newly synthesized DNA molecule with complementary sequence following the Watson-Crick complementary base pairing rule\cite{kornberg1992dna}. The movement of DNAP on the template occurs in stochastic deoxy-nucleotide triphospate (dNTP) base steps. One single base step of the DNAP corresponds to a single-nucleotide elongation of the new DNA strand. 

DNAP acts as a dual-purpose enzyme having polymerase as well as exonuclease activity during the replication process \cite{xie2009possible,sharma2012error,reha2010dna,johnson2010kinetic}. While its primary function is to elongate the newly formed DNA strand, it can also  cleave the newly formed DNA strand and remove the last nucleotide when this has been transfered to the exonuclease site of the enzyme. This activity functions as a proofreading mechanism:   
Upon misincorporation (incorporation of a wrong dNTP), the DNAP switches its functionality and the new strand is transferred to the exonuclease site where the wrong nucleotide is cleaved from the new strand through hydrolysis \cite{kunkel2000dna}, before the DNA strand returns to the active site for polymerization for elongation to continue \cite{reha2010dna}. This mechanism represents an intrinsic proof-reading mechanism known as exonucleolytic proofreading \cite{bkebenek2018fidelity,benkovic2001replisome,kunkel2004dna,kornberg1992dna,ibarra2009proofreading,kunkel2000dna} as misincorporated nucleotides are excised, thus resetting the incorporation process, such that the errors can be corrected by a second attempt at correct incorporation. Polymerase and exonuclease sites of the DNAP are separated from each other by a distance of 3-4
nm~\cite{ibarra2009proofreading,freemont1988cocrystal,wang1997crystal,kamtekar2004insights}, and the transfer of the new DNA strand between these two sites may involve one or many
intermediates~\cite{shamoo1999building,hogg2004crystallographic,subuddhi2008use}. The dynamics of the transfer reaction is not understood in detail. 

Just like wrong nucleotides can erroneously get incorporated during polymerization, compromising the fidelity of replication \cite{fersht1982kinetic}, correctly incorporated nucleotides can erroneously get transferred to the exonuclease site and be cleaved, resetting the incorporation process to the start and hence compromising the speed of elongation.. 
Thus, the coordinated action of the polymerase and exonuclease activity of DNAP is crucial for an effective and faithful replication mechanism and determined the velocity and accuracy are the two fundamental parameters in deciding the speed as well as the fidelity of replication. 

It is usually argued that an enhancement of accuracy results in a slow down of the elongation process, resulting in a  speed-accuracy trade-off \cite{johansson2012genetic,johansson2008rate,
lovmar2006rate,wohlgemuth2011evolutionary}. The high fidelity or enhancement of accuracy is believed to be as a matter of proof-reading mechanism during the exonuclease activity \cite{bkebenek2018fidelity,murugan2014discriminatory}. 
However, a recent study \cite{banerjee2017elucidating} claims that the speed-accuracy trade-off is not universal, but rather depends on the kinetic parameters of the enzyme. 
In this work we study a simple kinetic model of DNA replication, for which we can exactly calculate the velocity and accuracy of replication to investigate the interplay of the two functions of DNAP. Based on that analysis we compare different possible schemes of discrimination between correct and incorrect nucleotides and address how 
an accurate replication mechanism can be achieved, while also limiting the negative effect on the speed of elongation.
\begin{figure*}[htb!]
\centering
\includegraphics[width=16cm,height=6cm]{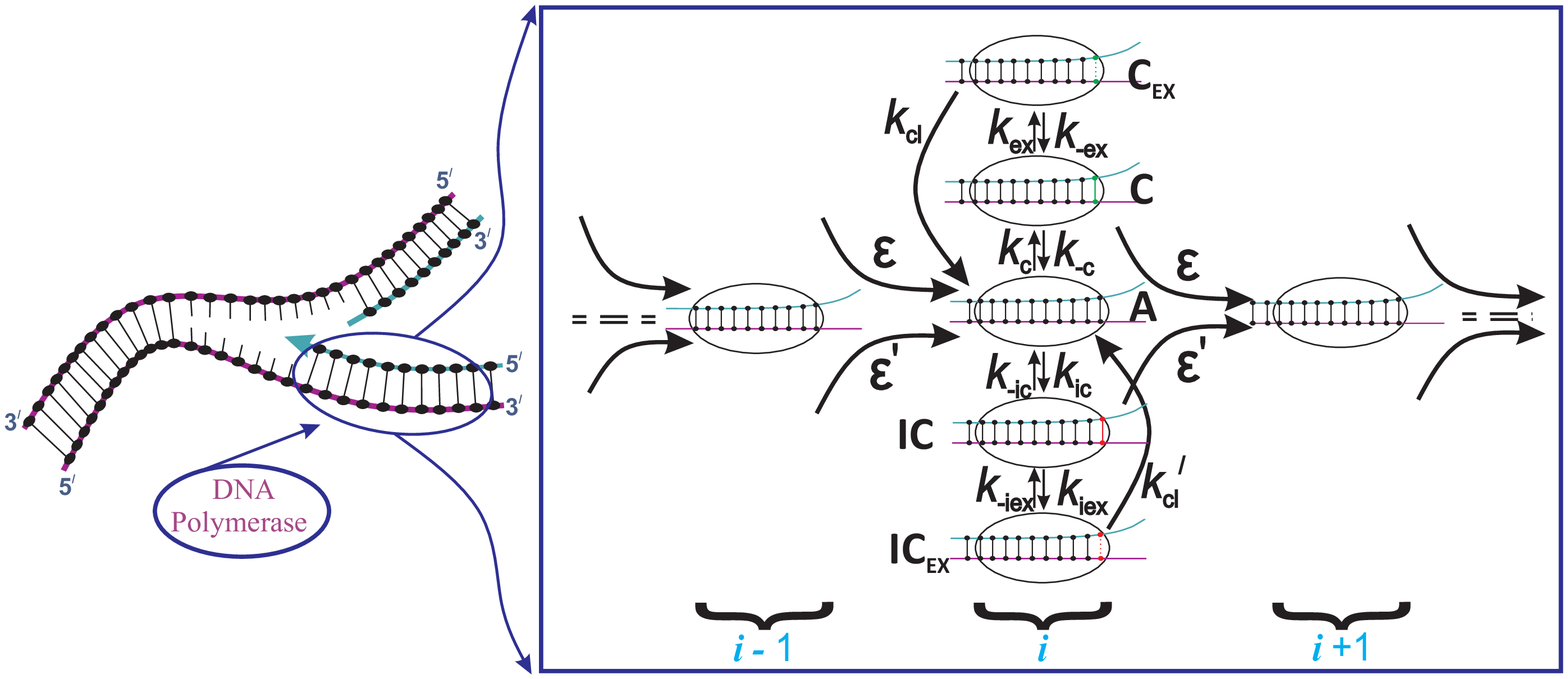}
\caption{Minimal model for the dynamics of DNA polymerase (DNAP). Left panel: Schematic depiction of the replication process with the leading DNA strand passing  through the DNA polymerase (encircled area). Right panel: Kinetic model for DNA replication. Three sites on the DNA are labelled as $i-1$, $i$, and $i+1$ and for the $i^{\mathrm{th}}$ site, the states of the DNAP are labelled as $A$, $C$, $IC$, $C_{EX}$, and $IC_{EX}$, respectively. Binding of a correct nucleotide takes the DNAP to the state $C$, whereas binding of an incorrect nucleotide takes the DNAP to the state $IC$. Actively elongating DNAP takes a forward step with rate $\epsilon$ from the state $C$ and with rate $\epsilon^{'}$ from the state $IC$. Exonuclease activity is described by the states $C_{EX}$ and $IC_{EX}$, depending on whether it is initiated from a correct or from an incorrect nucleotide. The arrows indicate the transition from one state to the other state with their corresponding transition rates as mentioned. The bond marked in red in state $IC$ corresponds to the nucleotide mismatch and thus to a weaker bond in state-$IC_{EX}$.
}
\label{mechanism}
\end{figure*}

\section{\textbf{Model}}
\subsection{Model with parallel pathways for correct and incorrect dNTP incorporation}

To study the interplay of polymerization and exonuclease activity, we use a minimal stochastic model, which is depicted in Figure~\ref{mechanism} as a  schematic diagram. The DNA template is considered as an one-dimensional lattice of certain length and each site of that lattice represents an individual base of the DNA template. The DNA polymerase (DNAP) moves on the DNA template by single base steps, simultaneously extending the copy by one deoxy-nucleotide triphosphate (dNTP) in each step. As a minimal description, we describe this process by Michaelis-Menten kinetics, reversible binding of the dNTP followed by irreversible incorporation into the new DNA strand. We note that this kinetics have been characterized in more detail and exhibit additional substeps, but here we aim at a minimal model. In the stochastic model, these two substeps are represented by transitions from the active state with a free binding site ($A$) to the occupied state (denoted by $C$) and back, where there are two pathways for the enzyme to transition back to the free state, either by unbinding of the dNTP or by incorporation and stepping, in which case the active and free state is reached at the subsequent position and with a one based longer new DNA strand. The rates of the three transitions are denoted as $k_c$, $k_{-c}$ and $\epsilon$, respectively.

To account for misincorporations and thus for errors in replication, the model includes a second pathway with Michaelis-Menten kinetics that describes exactly the same steps as the previous one, however for binding and incorporation of an incorrect, i.e. non-cognate dNTP. The corresponding occupied state is denoted by $IC$ and the rates, which may generally differ from those in the correct pathway, with $k_{ic}$, $k_{-ic}$ and $\epsilon'$, respectively.

\subsection{Proofreading pathway}
This basic model is extended to include the exonuclease activity of DNA polymerase. For that, the polymerase transitions to a state with the newly added dNTP in the exonuclease site.
The exonuclease state is denoted by $C_{EX}$ or state-$C_{IEX}$, depending on whether the new nucleotide is a correct or an incorrect one. The transition to that state occurs from state-\textit{C} (or state-\textit{IC}) with rate $k_{\mathrm{ex}}$ (or $k_{\mathrm{iex}}$), respectively. The reverse transition is also included in the model (with rates $k_{\mathrm{-ex}}$ and $k_{\mathrm{-iex}}$). Finally, the exonuclease reaction occurs from state $C_{EX}$ or state-$C_{IEX}$ with the cleavage rate $k_{\mathrm{cl}}$ or $k_{\mathrm{cl}}^{'}$, respectively. In this transition, the last incorporated nucleotide is cleaved off and the enzyme  goes back to the free state $A$. 

We note that the reaction scheme described here differs from the classical kinetic proofreading scheme \cite{hopfield1974kinetic}, as the proofreading step requires a transition away from the main pathway, namely the transfer of the new DNA strand from the polymerization site to the exonuclease site.
A reaction scheme that is very similar to ours (and also describes the off-pathway proofreading step) has been used by Sharma and Chowdhury \cite{sharma2012error}. Their focus however was on the stochasticity of DNA elongation by calculating dwell time distributions rather than on the different mechanisms of discrimination between the correct and incorrect nucleotides and on how they affect the design of accurate replication and the speed-accuracy trade-off. There are also some smaller differences between these two models. In their model, they include a transition to the exonuclease site before incorporation, which does not affect the accuracy (and is not included here), and do not explicitly represent the correct and incorrect incorporations as parallel pathways.  We also note that a  related, but considerably more complex proofreading scheme is found in RNA polymerase: RNA polymerase transitions to an off-pathway state for proofreading, the backtracking state, where it however performs diffusive motion and can cleave longer pieces of the newly synthesized RNA \cite{erie1993,sahoo2013backtracking}.

For discrimination between correct and incorrect nucleotides, we generally expect the following relations to hold: In the initial binding step, binding of incorrect nucleotides should be suppressed compared to binding of correct nuclotides. This may be due to a lower binding rate and/or a higher unbinding rate, reflecting a higher affinity for correct than for incorrect binding. The discrimination achieved in this step is restricted thermodynamically, by the difference in binding energy between the correct and incorrect dNTPs \cite{hopfield1974kinetic}. Additional discrimination can be expected based on the irreversible step (translocation of the enzyme), typically with $\epsilon'<\epsilon$. In contrast to binding, discrimination by the translocation rate is kinetic rather than thermodynamic and thus is not restricted by the binding energies but rather by the difference in the rates or in the corresponding energy barriers. Finally, the proofreading pathway is expected to also contribute kinetically to the accuracy, as transfer to the exonuclease site competes kinetically with the translocation of the enzyme. If the transfer for incorrect nucleotides is faster than that for correct nucleotides (i.e., for $k_{\mathrm iex}> k_{\mathrm ex}$), while at the same time if the translocation is slower ($\epsilon'<\epsilon$), one can expect the typical fate of an incorrect nucleotide to be excision and that of a correct one translocation. This tendency will be further enhanced if for correct nucleotides in the exonucleotide site, return to the polymerization site is more likely than cleavage, $k_{\mathrm{-ex}} > k_{\mathrm{cl}}$, while it is just the opposite for the incorrect nucleotides in the exonuclease site, $k_{\mathrm{-iex}} < k_{\mathrm{cl}'}$.  Estimates of the rates from the literature are listed in Table I. Note that in some cases, these are estimated from more detailed kinetic schemes. All rates in this table are scaled with respect to the forward stepping rate $\epsilon$.

\begin{table*}[t!]
\caption{Estimated values of the model parameters}
\centering
\small\addtolength{\tabcolsep}{12pt}
\begin{tabular}{lc c c c}
\hline \hline
Parameter & Symbol & Value &Refs\\[0.3ex]
\hline
Forward stepping rate& $\epsilon$ & $300 s^{-1}$ & \onlinecite{tsai2006new,johnson1993conformational,xie2009possible,patel1991pre}\\
Binding rate (correct dNTP) & $k_{c}$ & $10^{-1} \epsilon$ & \onlinecite{johnson1993conformational}\\
Binding rate (incorrect dNTP) & $k_{ic}$ & $10^{-2}-10^{-1}\epsilon$ & \onlinecite{johnson2010kinetic}\\
Unbinding rate (correct dNTP) & $k_{-c}$ & $10^{-2}\epsilon$ & \onlinecite{johnson2010kinetic}\\
Unbinding rate (incorrect dNTP) & $k_{-ic}$ & $10^{-2}-10^{-1}\epsilon$ & \onlinecite{johnson2010kinetic,hopfield1974kinetic}\\
Erroneous stepping rate &$\epsilon^{'}$ & $10^{-3}-10^{-2}\varepsilon$ & \onlinecite{tsai2006new,johnson1993conformational,patel1991pre}\\
Exonuclease site transfer rate (correct nucleotide) & $k_{ex}$ & $10^{-3}\varepsilon$ & \onlinecite{johnson1993conformational}\\
Reverse transfer rate (correct nucleotide) & $k_{-ex}$ & $< 700 s^{-1}$ & \onlinecite{tsai2006new}, further comments\\
Exonuclease site transfer rate (incorrect nucleotide) & $k_{iex}$ & $10*k_{ex}$ & \onlinecite{johnson1993conformational}\\
Reverse transfer rate (incorrect nucleotide)  & $k_{-iex}$ & $10^{-4}\varepsilon$ & \onlinecite{johnson1993conformational}\\
Cleavage rate (correct nucleotide) & $k_{cl}$ & $10^{-2}\varepsilon$ & \onlinecite{johnson1993conformational}\\
Cleavage rate (incorrect nucleotide) & $k_{cl}^{'}$ & $10^{-1}\varepsilon$ & \onlinecite{hopfield1974kinetic}\\
\hline
\end{tabular}
\label{table:para}
\end{table*}

\subsection{Steady state solution}
In the following, we will  analyze the elongation speed and the accuracy of replication for two cases. In the first case, we consider a purely polymerizing scenario, where the DNA polymerase does not undergo the transition to the exonuclease state (i.e., $k_{\mathrm{ex}} = k_{\mathrm{iex}} = 0$). In the second case, we include the exonuclease activity. 
In both cases, we calculate the accuracy and velocity of elongation from the steady-state fluxes between the states of the discrete stochastic model described so far. 
The approach is similar to that used in our earlier work for backtracking of RNA polymerase \cite{sahoo2013backtracking} and provides the mean values of these quantities. Consideration of the fluctuations as in Ref. \cite{sharma2012error} requires time dependent methods. 
Effectively, our system is a stochastic 5-state model. 
In the construction of our model, we have assumed that there is no sequence dependence of the rates \cite{yager1991thermodynamic,bai2004sequence,tadigotla2006thermodynamic}, so  all the rates considered here should be considered averages over a possible sequence-dependence. To determine steady-state fluxes, we solve the five state model shown in Fig.~1 and first determine the steady state probabilities of the five different states of the DNAP. These probabilities are denoted by $P_{\mathrm{i}}$, where $i= \mathrm{A}$ (active state), $\mathrm{C}$ (correct state), $\mathrm{IC}$ (incorrect state), $\mathrm{C_{\mathrm{EX}}}$ (correct exonuclease state), and  $\mathrm{I_{\mathrm{CEX}}}$ (incorrect exonuclease state), respectively. We calculate the steady state probabilities by equating the total incoming flux and the total outgoing flux for each state, which can be expressed as
\begin{equation}
k_{\mathrm{c}}P_{\mathrm{A}} + k_{\mathrm{-ex}}P_{\mathrm{C_{\mathrm{EX}}}} - \epsilon P_{\mathrm{C}} - k_{\mathrm{ex}}P_{\mathrm{C}} - k_{\mathrm{-c}}P_{\mathrm{C}} = 0,
\label{fluxCorrect}
\end{equation}

\begin{equation}
k_{\mathrm{ic}}P_{\mathrm{A}} + k_{\mathrm{-iex}}P_{\mathrm{I_{\mathrm{CEX}}}} - \epsilon^{'}P_{\mathrm{I_{\mathrm{C}}}} - k_{\mathrm{iex}}P_{\mathrm{I_{\mathrm{C}}}} - k_{\mathrm{-ic}}P_{\mathrm{I_{\mathrm{C}}}} = 0,
\label{fluxIncorrect}
\end{equation}

\begin{equation}
k_{\mathrm{ex}}P_{\mathrm{C}} - k_{\mathrm{cl}}P_{\mathrm{C_{\mathrm{EX}}}} - k_{\mathrm{-ex}}P_{\mathrm{C_{\mathrm{EX}}}} = 0,
\label{fluxCorrectExo}
\end{equation}
and 

\begin{equation}
k_{\mathrm{iex}}P_{\mathrm{I_{\mathrm{C}}}} - k_{\mathrm{cl}}^{'}P_{\mathrm{I_{\mathrm{CEX}}}} - k_{\mathrm{-iex}}P_{\mathrm{I_{\mathrm{CEX}}}} = 0.
\label{fluxincorrectExo}
\end{equation}
Together with the normalization condition for the probabilities,  
\begin{equation}
P_{\mathrm{A}} + P_{\mathrm{C}} + P_{\mathrm{I_{\mathrm{C}}}} + P_{\mathrm{C_{\mathrm{EX}}}} + P_{\mathrm{I_{\mathrm{CEX}}}} = 1,
\label{fluxActive} 
\end{equation}
the steady state probability of all states can be found. The explicit expressions are included  in Appendix-A 

\section{\textbf{Results and Discussions}}
Using the steady state probabilities of individual states, the velocity of elongation ($V$) can be calculated as
\begin{equation}
\centering
V = \epsilon^{'}P_{\mathrm{I_{\mathrm{C}}}} +\epsilon P_{\mathrm{C}}.
\label{fullvelocity}
\end{equation}
Here, $\epsilon^{'}P_{\mathrm{I_{\mathrm{C}}}}$ and $\epsilon P_{\mathrm{C}}$ represent the contributions to the elongation velocity by the forward stepping to the next site after correct incorporation with stepping rate $\epsilon$ and after incorrect incorporation with rate $\epsilon^{'}$ respectively.
Plugging $P_{IC}$ and $P_{C}$ into Eqn.~(6), an explicit expression for $V$ can be obtained, which is given in Appendix-A.

\noindent Further, the accuracy($A$) of DNA replication can be calculated as the ratio of the correct flux and the total flux as
\begin{equation}
\centering
A = \dfrac{\epsilon P_{\mathrm{C}}}{\epsilon P_{\mathrm{C}} + \epsilon^{'}P_{\mathrm{I_{\mathrm{C}}}}}= \frac{1}{1+(\frac{\epsilon'}{\epsilon})\frac{P_{IC}}{P_{C}}}.
\label{fullaccuracy}
\end{equation}
By substituting the expression for the steady state probabilities, ($P_{\mathrm{C}}$ and $P_{\mathrm{IC}}$) in the above expression, we can obtain the accuracy $A$ of replication, which is given in the Appendix-A.

\subsection{Purely polymerizing Scenario}

To analyze the discrimination between correct and incorrect nucleotide and the effect of errors in the velocity, we first consider the purely polymerizing scenario, as observable in mutants without exonuclease activity. To that end, we set $k_{ex}=k_{iex}=0$, i.e. the DNA polymerase does not make a transition to the exonuclease state, neither from the correct binding state nor from the incorrect binding state. For this case, the elongation velocity $V_{pp}$ can be calculated as
\begin{equation} \label{eq1}
\begin{split}
V_{pp}
&=\dfrac{\varepsilon k_{c}(\varepsilon^{'}+k_{-ic})+\varepsilon^{'}k_{ic}(\varepsilon+k_{-c})}{(\varepsilon+k_{-c})(\varepsilon^{'}+k_{ic}+k_{-ic})+k_{c}(\varepsilon^{'}+k_{-ic})} 
\end{split}
\end{equation}

Similarly, the accuracy  $A_{pp}$ of replication can be calculated as
\begin{equation} \label{eq1}
\begin{split}
A_{pp}
&=\dfrac{\varepsilon k_{c}(\varepsilon^{'}+k_{-ic})}{\varepsilon\varepsilon^{'}k_{c}+\varepsilon \varepsilon^{'}k_{ic}+\varepsilon k_{c}k_{-ic}+\varepsilon^{'}k_{ic}k_{-c}}
\end{split}
\end{equation}

\begin{figure}[h]
\includegraphics[width=9cm,height=7cm]{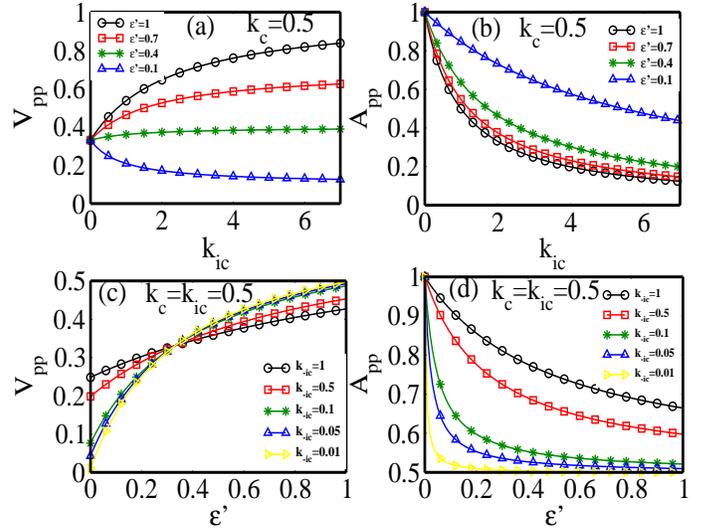}
\caption{Velocity $V_{\mathrm{PP}}$ and accuracy  $A_{\mathrm{PP}}$ in the purely polymerizing scenario, both (a,b) as a function of the binding rate of incorrect nucleotides, $k_{ic}$ and (c,d) as function of the stepping rate $\epsilon^{'}$. Other parameters are varied as indicated in the legends, the parameters $\epsilon=1$ and $k_{-c}=0.01$ are fixed.}
\label{Vvarykminuseprime}
\end{figure}

Figure 2 (a) and (b) shows the plot of velocity $V_{pp}$ and the accuracy $A_{pp}$ for a constant rate of correct nucleotide binding, varying the rate of incorrect nucleotide binding ($k_{ic}$). It is observed that the possibility of errors obviously decreases the accuracy whereas the influence of such errors may increase or decrease the velocity of elongation. $V_{pp}$ may increase or decrease with $k_{ic}$, depending on the value of stepping rate $\epsilon^{'}$ after an erroneous incorporation. Similarly Fig.~2(c) and Fig.~2(d) shows $V_{pp}$ and $A_{pp}$ as a function of $\epsilon^{'}$ for different values of $k_{-ic}$.  $V_{pp}$ has a finite value at $\epsilon^{'}=0$ and for any fixed value of $k_{-ic}$. This nonzero value of $V_{pp}$ is due to the forward stepping of the DNAP from the correct state to the next site with rate $\epsilon$. For $\epsilon^{'}=0$, the DNAP usually goes back to the active state for a nonzero value of $k_{-ic}$ and eventually incorporation of a correct nucleotide (and stepping) takes place. 
Thus for small $\epsilon'$, the typical fate of an incorrect nucleotide is to unbind, making the binding of an incorrect nucleotide a dead-end branch off the main pathway and rapid unbinding beneficial for high speed. By contrast, for large $\epsilon'$, incorrect nucleotides are incorporated and contribute to the velocity of polymerization. Moreover, interestingly we observe that for a critical value of $\epsilon^{'}$, (i.e., at $\epsilon^{'}=\epsilon_{c}^{'}$), at which the transition between these two regimes occurs, the velocity is independent of $k_{-ic}$. The accuracy (Fig. 2(d)), on the other hand, always increases when the unbinding rate is increased, independent of the value of $\epsilon'$. Just like decreasing the binding rate of incorrect nucleotides, increasing the unbinding rate  $k_{-ic}$ always increases the accuracy, but may increase or decrease the velocity.

The critical stepping rate $\epsilon_{c}^{'}$ at which $V_{pp}$ is independent of $k_{-ic}$ is exactly calculated as  $\epsilon_{c}^{'}=\frac{\epsilon k_{c}}{\epsilon+k_{c}+k_{-c}}$ when $k_{c}=k_{ic}$. From this expression one can notice that $\epsilon_{c}^{'}$ increases as a function of $k_{c}$ and for large $k_{c}$ ($k_{c} \rightarrow \infty$), it approaches $\epsilon$, as plotted in Fig.~3. This implies that $\epsilon_{c}^{'}$ can be controlled by fine tuning the $k_{c}$ values. 

\begin{figure}[htb!]
\centering
\includegraphics[width=8.6cm,height=7cm]{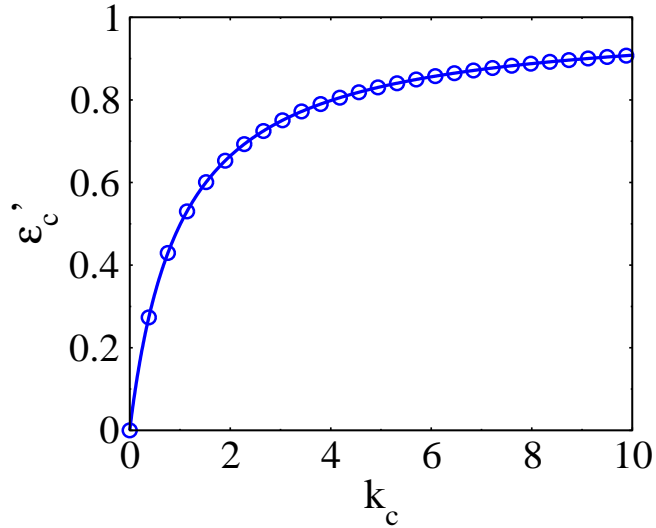}
\caption{Critical stepping rate $\epsilon_{c}^{'}$ as a function of the binding rate $k_{c}$. The fixed parameters are $\epsilon=1$ and $k_{-c}=0.01$.}
\label{nochange}
\end{figure}

The scenario considered so far, is the simplest exonuclease-deficient scenario. Another exonuclease-deficient and therefore purely polymerizing scenario is when transfer to the exonuclease site is possible, but with no cleavage  (i.e., with $k_{cl}=k_{cl}^{'}=0$). Velocity and accuracy in this case are plotted as a function of $k_{ic}$ together with the corresponding results for the case without transfer to the exonuclease site (i.e., for $k_{ex}=k_{iex}=0$) in Fig.~4. The velocity is generally reduced by the transfer to the (inactive) exonuclease site, whereas the accuracy is seen to be the same in both scenarios.

\begin{figure}[h]
\includegraphics[width=9cm,height=7cm]{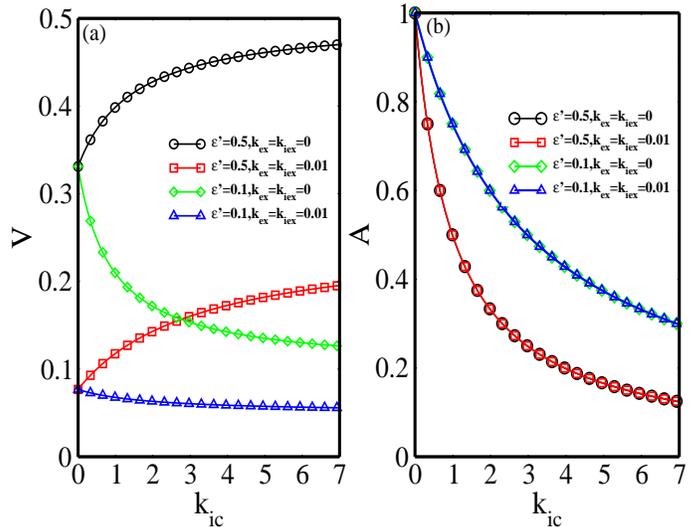}
\caption{Sceanario with transfer to the exonuclease site, but without exonuclease activity: Velocity $V$ and accuracy $A$ as a function of the binding rate $k_{ic}$ of incorrect nucleotides for different values of the stepping rate $\epsilon^{'}$ and the exonuclase site transfer rates $k_{ex}$ and $k_{iex}$. 
The other parameters are fixed:  $k_{c}=k_{ic}=0.5$, $\epsilon=1$, $k_{-c}=k_{-iex}=0.01$, and $k_{-ex}=0.001$.}
\label{Vvarykminuseprime}
\end{figure}

\subsection{Polymerization with exonuclease activity}
\begin{figure}[htb!]
	\begin{center}
		\includegraphics[width=9cm, height=8cm]{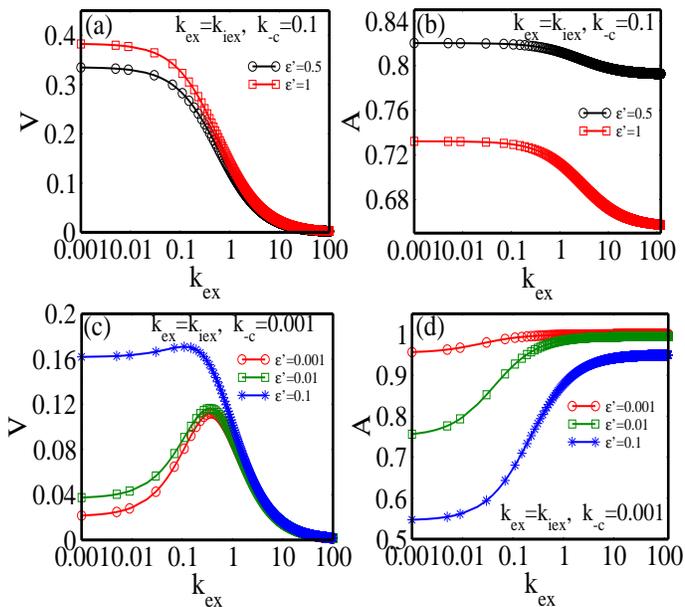}
		\caption{Variety of behaviors in a scenario with exonuclease activity: Velocity $V$ and accuracy $A$ as a function of the exonuclease transfer rate $k_{ex}$  for two different values of $k_{-c}$ [in (a,b) and in (c,d), respectively]. In all cases, there is no discrimination between correct and incorrect nucleotides in the exonuclease pathway. The   parameters are $k_{c}=k_{ic}=0.5$, $\epsilon=1$, $ k_{\mathrm{cl}} = k_{\mathrm{-ex}} = k_{\mathrm{-iex}} = 0.1 $, and $ k_{\mathrm{cl}}^{'} $ = $k_{cl}*\exp(3)$. $ k_{\mathrm{-ic}}= k_{\mathrm{-c}}*\exp(3)$. Note that both increase and decrease of velocity and accuracy are seen and that the two are not generally anticorrelated. }
		\label{VelocityAccuracy}
	\end{center}
	
\end{figure}

\begin{figure}[htb!]
	\begin{center}
		\includegraphics[width=9cm, height=6cm]{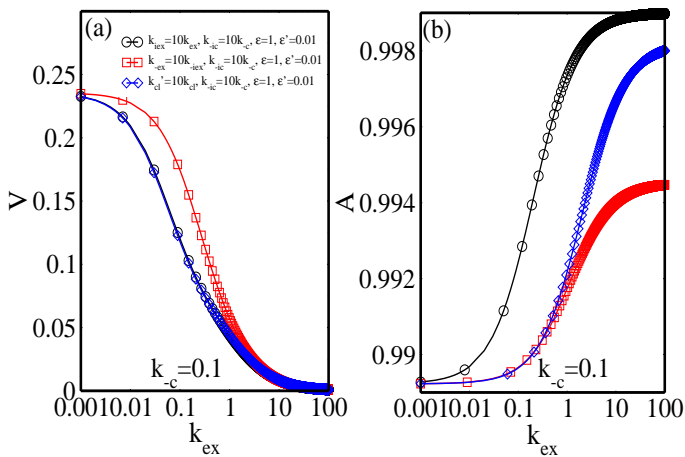}
\caption{Three scenarios with discrimination between correct and incorrect nucleotides in the exonuclease pathway: (a) Velocity $V$ and (b) accuracy $A$ as a function of $k_{ex}$. The three cases depict discrimination in the exonuclease transfer rate, the reverse transfer rate and the cleavage rate, respectively. The fixed parameters are $k_{c}=k_{ic}=0.5$.}
\label{VelocityAccuracy}
	\end{center}
\end{figure}

Including the exonuclease activity, we have exactly calculated the elongation velocity, $V$ as well as the accuracy of replication, $A$ using Eq.~(16) and Eq.~(19), respectively. We expect that the typical effect of a proofreading mechanism is a reduction of the velocity and an increase in accuracy. However, the following consideration shows that this is not generally the case: Without discrimination  between the correct and incorrect nucleotides in the exonuclease transfer step ($k_{ex}=k_{iex}$), we observed that both velocity as well as accuracy decrease with increasing $k_{ex}$ [Fig.~5(a) and 5(b)]. The decrease in accuracy can be understood as follows: Once a nucleotide, correct or incorrect, is bound, it can either be incorporated irreversibly by the forward step or be removed again by one of two pathways, unbinding or exonuclease transfer and cleavage. If the cleavage rate is sufficiently high, then the cleavage follows almost deterministically upon exonuclease transfer, and the effective rate for the removal of the nucleotide is $k_{-c}+k_{ex}$ (or $k_{-ic}+k_{iex}$). If the first term discriminates between the correct and the incorrect nucleotides, but the second does not, increasing the second will inevitably reduce the accuracy rather than increasing it, as expected for a proofreading pathway. Thus, additional discrimination in the exonuclease pathway is required for proofreading. 

We also note that the decrease of the velocity with increasing $k_{ex}$ is also not universal. When there is no discrimination between the correct and incorrect nucleotides in the exonuclease transfer step, for very small value of $\epsilon^{'}$,  the velocity initially increases with $k_{ex}$ for very small values of $k_{-c}$, shows a maximum and then decreases with $k_{ex}$ [Fig.~5(c)]. Whereas the accuracy increases with $k_{ex}$ [Fig.~5(d)]. The increase of velocity can be understood as follows: The presence of incorrect nucleotide reduces the velocity due to the corresponding slow stepping. Exonuclease can circumvent that slow stepping. However for high exonuclease transfer rates, the velocity is decreased as exonuclease transfer now also successfully compete with the rapid stepping after correct incorporation.
The presence of a maximum in the velocity is understood as preventing slow stepping after incorrect nucleotide binding, but not rapid stepping after correct binding, suggests one principle how the exonuclease transfer rate should be set, namely between the two stepping rates (see below).

These results shed some light on the question of a  trade-off between velocity and accuracy, which is generally expected for proofreading mechanisms and has been reported in a number of studies \cite{wohlgemuth2011evolutionary,johansson2012genetic,johansson2008rate}. Our results show that such trade-off is also seen in our minimal model, provided the kinetic parameters are in the range expected for DNA polymerases. The observation of a trade-off is  independent of the mechanisms of discrimination in the proofreading pathway as discussed below. 
However, Figure 5 shows that  speed-accuracy tradeoff is  not universal, consistent with a recent study \cite{banerjee2017elucidating}, but rather dependent on the kinetic model parameters. This is a typical feature of kinetic proofreading mechanism and can be due to the discrimination of free energy associated with different reactions. 

Since additional discrimination between correct and incorrect nucleotides is needed for the exonuclease activity to result in an increase in accuracy, we next compared the three possibilities for discrimination in exonuclease pathway: Using realistic parameters for the polymerization pathway, including discrimination in unbinding ($k_{-ic}>k_{-c}$) and in stepping ($\epsilon'<\epsilon$), we determined the velocity and the accuracy for (i) $k_{iex}>k_{ex}$, (ii) $k_{-iex}<k_{-ex}$, and for (iii) $k_{cl}'>k_{cl}$. In all these three different cases, all other rates of the exonuclease pathway are the same for correct and incorrect nucleotides, and we take the rates that are different to differ by a factor of 10. The latter factor is of the order of $e^{\frac{\Delta G}{k_BT}}$ with  a mismatch energy $\Delta G\sim 3 k_{B} T$.

Fig.~6 shows the comparison between the three cases. In all cases, the accuracy increases with increasing exonuclease transfer rate, while the velocity decreases.
However there are some marked differences between the scenarios: Discrimination in the transfer rate or in the cleavage rate show an almost identical decrease in the velocity and a very similar increase in accuracy. However, the increase in accuracy is shifted towards higher values of $k_{ex}$ for the case of discrimination in the transfer rate as compared to the discrimination in the cleavage rate. Thus, for the same transfer rate (and for the same decrease in velocity), the discrimination in transfer results in a higher accuracy than discrimination in cleavage. Discrimination in the reverse transfer rate, by comparison, shows an overall smaller increase in accuracy as well as smaller decrease in velocity. 

\section{Conclusions}
In this paper, we have studied a simple kinetic model of DNA replication and have exactly calculated the accuracy and elongation velocity in the absence as well as in the presence of exonuclease activity. In general, the fidelity of DNA replication is based on the nucleotide selectivity of the DNA polymerase during incorporation, its exonucleolytic proofreading activity and the postreplicative DNA mismatch repair. The latter contribution is not included in our model, in which an error remains uncorrected once the new DNA strand is elongated and DNA polymerase has made a forward step. 
During the exonucleolytic proofreading, an erroneously incorporated nucleotide is cleaved off, hence the exonuclase activity improves the overall accuracy of replication, but at the same time it also typically reduce the speed of elongation by "resetting" the replication process to the beginning of the last step. Similarly the escape of errors (via erroneous stepping) speeds up the elongation mechanism but compromises the overall accuracy. The tradeoff between speed and accuracy however is not universal, but depends on the kinetic parameters, in agreement with the earlier results \cite{banerjee2017elucidating}. Our minimal model provides a simple picture of the design of an accurate replication system. Through the modulation of the  model parameters, we can identify several kinetic principles. Discrimination between the correct and incorrect nucleotides arises both from the polymerization pathway as well as from the proofreading pathway and is based on the kinetic competition in both the cases. For both, slow forward stepping after incorporation of an incorrect nucleotide (slower than for a correct nucleotide) is crucial for the accuracy, giving time for unbinding and transfer to the exonuclease site, respectively. However, additional discrimination in the exonuclease pathway is needed and provided most efficiently by the transfer reaction. Thus the accuracy depends on the  coordinated action of the polymerase and exonuclease activity of DNA polymerase. 

\section{Acknowledgments}
MS acknowledges the INSPIRE Faculty award (IFA 13 PH-66) by the Department of Science and Technology, India and the Faculty Recharge Program (FRP-56055) of UGC, India for the financial support. 

\section{Appendix}
In this appendix, we give the explicit expressions for the steady-state probabilities obtained as the solution to Eqs. (1)-(5):
\begin{equation}
P_{\mathrm{C}} = k_{\mathrm{c}}(k_{\mathrm{-ex}} + k_{\mathrm{cl}})\{k_{\mathrm{iex}}k_{\mathrm{cl}}^{'} + (k_{\mathrm{-iex}} + k_{\mathrm{cl}}^{'})(\epsilon^{'} + k_{\mathrm{-ic}})\}/T,
\label{probabilityCorrect}
\end{equation}

\begin{equation}
P_{\mathrm{I_{\mathrm{C}}}} = k_{\mathrm{ic}}(k_{\mathrm{-iex}} + k_{\mathrm{cl}}^{'})\{k_{\mathrm{ex}}k_{\mathrm{cl}} + (k_{\mathrm{-ex}} + k_{\mathrm{cl}})(\epsilon + k_{\mathrm{-c}})\}/T,
\label{probabilityError}
\end{equation}

\begin{equation}
P_{\mathrm{C_{\mathrm{EX}}}} = k_{\mathrm{ex}}k_{\mathrm{c}}\{k_{\mathrm{iex}}k_{\mathrm{cl}}^{'} + (k_{\mathrm{-iex}} + k_{\mathrm{-cl}}^{'})(\epsilon^{'} + k_{\mathrm{-ic}})\}/T,
\label{probabilityB}
\end{equation}

\begin{equation}
P_{\mathrm{I_{\mathrm{CEX}}}} = k_{\mathrm{iex}}k_{\mathrm{ic}}\{k_{\mathrm{ex}}k_{\mathrm{cl}} + (k_{\mathrm{-ex}} + k_{\mathrm{cl}})(\epsilon + k_{\mathrm{-c}})\}/T,
\label{probabilityBprime}
\end{equation}
and
\begin{equation}
P_{\mathrm{A}}= 1 - P_{\mathrm{C_{\mathrm{EX}}}} - P_{\mathrm{I_{\mathrm{CEX}}}} - P_{\mathrm{I_{\mathrm{C}}}} - P_{\mathrm{C}}.
\label{probabilityActive}
\end{equation}

In the above expressions, $T$ is given by
\begin{equation}
\begin{split}
T = k_{\mathrm{iex}}\{k_{\mathrm{ex}}k_{\mathrm{cl}}^{'}k_{\mathrm{c}} + k_{\mathrm{ex}}k_{\mathrm{cl}}(k_{\mathrm{ic}} + k_{\mathrm{cl}}^{'}) +(k_{\mathrm{-ex}} + k_{\mathrm{cl}})\\
((k_{\mathrm{ic}} + k_{\mathrm{cl}}^{'})(\epsilon + k_{\mathrm{-c}}) + k_{\mathrm{cl}}^{'}k_{\mathrm{c}}) \} + (k_{\mathrm{-iex}} + k_{\mathrm{cl}}^{'}) \\
\{k_{\mathrm{ex}}k_{\mathrm{ic}}k_{\mathrm{cl}} + k_{\mathrm{ex}}(k_{\mathrm{c}} + k_{\mathrm{cl}})(\epsilon^{'} + k_{\mathrm{-ic}}) + (k_{\mathrm{-ex}} + k_{\mathrm{cl}})\\
(k_{\mathrm{c}}(\epsilon^{'} + k_{\mathrm{-ic}}) + (k_{\mathrm{ic}} + (\epsilon^{'} + k_{\mathrm{-ic}}))(\epsilon+ k_{\mathrm{-c}}))\}.
\end{split}
\end{equation}

We also give explicit expressions for the velocity and accuracy as obtained from Eqs. (6) and (7):
The velocity $V$ can be expressed as
\begin{equation}
V = \dfrac{S}{U}
\end{equation}
with $S$ and $U$ are given by
\begin{equation}
\begin{split}
S = \epsilon^{'}k_{\mathrm{ic}}(k_{\mathrm{cl}}^{'} + k_{\mathrm{-iex}})\{k_{\mathrm{cl}}k_{\mathrm{ex}} + (\epsilon+ k_{\mathrm{-c}})(k_{\mathrm{-ex}} + k_{\mathrm{cl}})\}\\
+ \epsilon k_{\mathrm{c}}(k_{\mathrm{cl}} + k_{\mathrm{-ex}})\{k_{\mathrm{cl}}^{'}k_{\mathrm{iex}} + (\epsilon^{'} + k_{\mathrm{-ic}})(k_{\mathrm{cl}}^{'} + k_{\mathrm{-iex}})\}\\
\end{split}
\end{equation}
and 
\begin{equation}
\begin{split}
U = (k_{\mathrm{cl}}^{'} + k_{\mathrm{-iex}})\{(k_{\mathrm{cl}} + k_{\mathrm{-ex}})((\epsilon + k_{\mathrm{-c}})(\epsilon^{'} + k_{\mathrm{ic}} + k_{\mathrm{-ic}})\\
+ k_{\mathrm{c}}(\epsilon^{'} + k_{\mathrm{-ic}})) + k_{\mathrm{ex}}(\epsilon^{'} + k_{\mathrm{-c}})(k_{\mathrm{c}} + k_{\mathrm{cl}}) + k_{\mathrm{ic}}k_{\mathrm{cl}}k_{\mathrm{ex}}\} \\
+ k_{\mathrm{iex}}\{(k_{\mathrm{cl}} + k_{\mathrm{-ex}})(k_{\mathrm{c}}k_{\mathrm{cl}}^{'} + (\epsilon+ k_{\mathrm{-c}})(k_{\mathrm{ic}} + k_{\mathrm{cl}}^{'}))\\
+ k_{\mathrm{cl}}k_{\mathrm{ex}}(k_{\mathrm{ic}} + k_{\mathrm{cl}}^{'}) + k_{\mathrm{c}}k_{\mathrm{cl}}^{'}k_{\mathrm{ex}}\}.
\end{split}
\end{equation}

The accuracy $A$ is found to be
\begin{equation}
A=\dfrac{1}{1+\dfrac{\epsilon^{'}}{\epsilon}N},
\end{equation}
where
\begin{equation}
N=\dfrac{k_{ic}(k_{-iex}+k_{cl}^{'})[k_{ex}k_{cl}+(k_{-ex}+k_{cl})(\epsilon+k_{-c})]}{k_{c}(k_{-ex}+k_{cl})[k_{iex}k_{cl}^{'}+(k_{-iex}+k_{cl}^{'})(\epsilon^{'}+k_{-ic})]}.
\end{equation}

\bibliographystyle{unsrt}
\nocite{*}

\end{document}